\documentclass[aps, preprint, nofootinbib,preprintnumbers,eqsecnum,superscriptaddress,sort]{revtex4}
\pdfoutput=1
\usepackage{amssymb,amsmath,amsfonts,amsthm,latexsym,mathrsfs}
\usepackage{graphics, graphicx,color,epsfig,subfigure}
\usepackage[small,hang,justification=raggedright,singlelinecheck=on]{caption}
\usepackage[
      colorlinks=true,
      linkcolor=blue,
      urlcolor=blue,
      filecolor=black,
      citecolor=red,
      pdfstartview=FitV,
      pdftitle={},
        pdfauthor={Gary Horowitz, Jorge Santos, Benson Way},
        pdfsubject={},
        pdfkeywords={},
        pdfpagemode=None,
        bookmarksopen=true
      ]{hyperref}
\numberwithin{equation}{section}
\def \be {\begin{equation}}
\def \ee {\end{equation}}
\def \bea {\begin{eqnarray}}
\def \eea {\end{eqnarray}}
\def \dd {\mathrm{d}}

\newcommand{\bln}{\begin{align}}
\newcommand{\eln}{\end{align}}
\newcommand{\bst}{\begin{split}}
\newcommand{\est}{\end{split}}
\newcommand{\bi}{\begin{itemize}}
\newcommand{\ei}{\end{itemize}}
\newcommand{\ben}{\begin{enumerate}}
\newcommand{\een}{\end{enumerate}}

\def\eeq{\end{equation}}

\begin{document}
\title {Evidence for an Electrifying Violation of Cosmic Censorship}

\author{Gary T. Horowitz}
\email{gary@physics.ucsb.edu}
\affiliation{Department of Physics, UCSB, Santa Barbara, CA 93106}

\author{Jorge E. Santos}
\email{jss55@cam.ac.uk}
\affiliation{Department of Applied Mathematics and Theoretical Physics, University of Cambridge, Wilberforce Road, Cambridge CB3 0WA, UK \vspace{1 cm}}

\author{Benson Way}
\email{bw356@cam.ac.uk}
\affiliation{Department of Applied Mathematics and Theoretical Physics, University of Cambridge, Wilberforce Road, Cambridge CB3 0WA, UK \vspace{1 cm}}

\begin{abstract}\noindent{
We present a plausible counterexample to cosmic censorship in four dimensional Einstein-Maxwell theory with asymptotically anti-de Sitter boundary conditions. Smooth initial data evolves to a region of arbitrarily large curvature that is visible to distant observers.  Our example is based on a holographic model of an electrically charged, localised defect which was previously studied at zero temperature.  We partially extend those results to nonzero temperatures. 
}
\end{abstract}
\maketitle

\tableofcontents

\section{Introduction and Summary}

Proving or finding a counterexample to cosmic censorship \cite{Penrose:1969pc} is probably the most important open problem in four dimensional classical general relativity. While various statements of cosmic censorship have been proposed, we will be interested in the possibility of forming a region of arbitrarily large curvature that is visible to distant observers. In higher dimensions, there is strong numerical evidence that this form of cosmic censorship fails \cite{Lehner:2011wc}, since black strings pinch off due to the Gregory-Laflamme instability \cite{Gregory:1993vy}. In four dimensions, this instability does not exist, and less is known.

Motivated by gauge/gravity duality, there has been considerable interest in general relativity with asymptotically anti-de Sitter (AdS) boundary conditions. A plausible candidate for a vacuum counterexample to cosmic censorship (with $\Lambda < 0$) has recently been proposed based on the superradiant instability of Kerr-AdS black holes \cite{Dias:2015rxy,Niehoff:2015oga}.  We will present evidence for another plausible counterexample using electric fields. That is, we restrict ourselves to solutions to the bulk action
\begin{equation}
S = \frac{1}{16\pi G}\int \mathrm{d}^4 x\,\sqrt{-g}\left[R+\frac{6}{L^2}-F^{ab}F_{ab}\right]\,,
\label{eq:action}
\end{equation}
where $L$ is the AdS length scale, and $F\equiv \dd A$ is the Maxwell field strength.

With AdS boundary conditions, one is free to specify the (conformal) boundary metric at asymptotic infinity, as well as the asymptotic form of the vector potential $A_a$. We choose the boundary metric to be flat (as in the standard Poincar\'e coordinates for AdS)  
\be\label{eq:bndisflat}
\dd s^2_\partial = -\dd t^2+\dd r^2+r^2 \dd\phi^2\;,
\ee
and the potential to asymptotically have only a nonzero time component
\be\label{eq:bndgaugefield}
A|_\partial =\mu(t,r,\phi)\mathrm{d}t\;.
\ee
We have recently studied this problem with $\mu=a\,p(r)$, where $a$ is a constant amplitude and $p(r)$ is a radial profile that vanishes at large radius: $ \lim_{r\to\infty}p(r)=0$ \cite{Horowitz:2014gva}.  While our current motivation for studying such systems stems from cosmic censorship, the motivation in \cite{Horowitz:2014gva} was to understand the effect of localised charged defects in a simple holographic model for condensed matter systems.

Let us briefly review some of the results of the analysis in \cite{Horowitz:2014gva}, which was restricted to static solutions at zero temperature. For profiles $p(r)$ that vanish faster than $1/r$ at large $r$, there are solutions with a standard Poincar\'e horizon in the interior.  One family of such solutions describe static, self-gravitating electric fields in AdS.  This family extends from $a=0$, where it meets with pure Poincar\'e AdS, to a maximum amplitude $a=a_{\max}$, where a naked curvature singularity appears.

We propose that this singularity allows for a violation of cosmic censorship. Consider the following dynamical scenario.  Set $\mu(t,r) = a(t) p(r)$ where $a$ is initially zero, and slowly increases to a constant value larger than $a_{\max}$. If the amplitude is increased sufficiently slowly, then the bulk solution is well-approximated by a slowly evolving family of static solutions. If the endpoint of such an evolution is the singular static solution, then cosmic censorship will be violated. It is certainly possible that naked singularities will not form in finite time, but the curvature on the horizon should grow without 
bound\footnote{In critical gravitational collapse, there is an open set of initial data where the curvature exceeds a prescribed bound. Our result is stronger since the curvature generically grows without any bound.}.

Let us address some obstacles to this proposal.  First, if a black hole forms, then the singularity may not be naked. Indeed, there are regular static solutions containing a `hovering' black hole.  These are spherical, extremal, Reissner-Nordstrom-AdS black holes that sit above the Poincare horizon.  The usual gravitational  attraction of this black hole towards the Poincar\'e horizon is balanced by an electrostatic attraction towards the boundary at infinity.  At zero temperature, these solutions appear for values of $a$ above a critical amplitude $a_*$ where $a_*<a_{\max}$, and seem to exist for arbitrarily large values of $a$ \cite{Horowitz:2014gva}.  However, hovering black holes cannot form in our dynamical process since there is no charged matter to collapse to form a charged black hole. 

The Poincar\'e horizon cannot be uniformly heated to a planar horizon that covers the singularity either since that would require infinite energy. Though the time-dependent boundary condition injects energy into the bulk, $\mu(r)$ vanishes rapidly enough at large $r$ to ensure that the total energy remains finite.  

Another possible loophole is the following. While it was shown in \cite{Horowitz:2014gva} that solutions without hovering black holes become singular as $a \to a_{\max}$, it was not shown that this singularity persists for larger amplitude.  If there are nonsingular solutions with $a>a_{\max}$ (without hovering black holes), then the above counterexample would not be generic; it would require fine-tuning of the boundary data. We will argue below that this does not happen by constructing nonzero temperature solutions for $a>a_{\max}$, and showing that the curvature appears to diverge as $T\to 0$.

Adding temperature suggests an alternative way to violate cosmic censorship. Suppose we start with the hovering black hole solution with $a > a_{\max}$ and increase the temperature, keeping $\mu$ fixed. When $T > 0$, both the planar horizon and the hovering black hole become nonextremal. At a certain critical temperature, the two horizons could merge. This is similar to the usual merging of two black holes and does not violate cosmic censorship. However, suppose that we reverse this process by starting at high temperature (and $a > a_{\max}$) with a single connected horizon, and reduce $T$. In order to recover the hovering black hole at low $T$, the horizon will have to bifurcate. This results in a naked singularity. 

The static $T > 0$ solutions we will construct suggest that this scenario is unlikely.  The horizon does not pinch off in the space of static solutions, except at $T=0$. Instead, as one lowers the temperature, the planar horizon develops a bulge and eventually resembles a `black mushroom'\footnote{ Vacuum solutions with similar horizon geometry can be constructed analytically using the AdS C-metric \cite{Chen:2015zoa}.}. The `cap' of the black mushroom remains connected to the planar horizon, and does not split into a hovering black hole.  Additionally, while it may be possible for these static black mushrooms to produce naked singularities by pinching off dynamically, this approach is perhaps less physical than increasing $\mu$. Temperature is a property of equilibrium configurations, and imposing a time dependent temperature is not a well defined boundary condition for dynamical, asymptotically AdS solutions.

It turns out that these nonzero temperature solutions are non-unique.  That is, there is more than one such solution with the same $a$ and $T$.  This type of nonuniqueness is common for black holes in global AdS, and has even been seen in studies of planar black holes in AdS \cite{Horowitz:2014gva,Santos:2014yja}. Since this nonuniqueness appears to be confined to $a<a_{\max}$, at very low nonzero temperatures, it does not affect our proposed counterexample to cosmic censorship.   However, it does affect the complexity of the phase diagram for charged defects.

The remainder of this article is organised as follows.  In the next section, we detail our numerical construction of these $T>0$ solutions.  In section III, we discuss the results of this calculation.  This includes a description of the black mushrooms, an extrapolation of the behaviour of these solutions to $T=0$, and a discussion of non-uniqueness and the implications for hovering black holes and the phase diagram of charged defects. Finally, we make a few concluding remarks in section IV. 

\section{Solutions with Nonzero Temperature }

In this section we describe how to construct static, nonzero temperature solutions to \eqref{eq:action} with our chosen boundary conditions. We will use these results in the next section to argue that $T > 0$ solutions do not have naked singularities, while $T=0$ solutions  
with $a > a_{\max}$ do. 
 
 The action \eqref{eq:action} yields the equations of motion
\be\label{eq:einsteinmaxwell}
G_{ab}\equiv R_{ab}+\frac{3}{L^2}g_{ab} - 2\left(F_{ac}F_{b\phantom{c}}^{\phantom{b}c}-\frac{1}{4}g_{ab}F^{cd}F_{cd}\right)=0\;,\qquad
\nabla_a F^{ab}=0\;. 
\ee
We are interested in static, axisymmetric solutions with a timelike Killing vector $\partial_t$ and an axisymmetric Killing vector $\partial_\phi$.  Our solutions will depend upon the remaining two coordinates (\emph{i.e.}, the problem is cohomogeneity two). We are therefore searching for regular solutions to \eqref{eq:einsteinmaxwell} satisfying the boundary conditions \eqref{eq:bndisflat} and
\be\label{eq:bndgaugefieldstatic}
A|_\partial =\mu(r)\mathrm{d}t=a\,p(r)\dd t\,,
\ee
where $a$ is the amplitude and we have chosen a particular profile $p(r)$ given by
\begin{equation}\label{eq:profile}
p(r) = \frac{1}{\left(1+\frac{r^2}{\sigma^2}\right)^4}\,,
\end{equation}
which is a positive function that monotonically decreases with increasing $r$.  This profile  has a characteristic length scale $\sigma$.  As seen in \cite{Horowitz:2014gva}, we do not expect the main features of our results to change for different choices of $p(r)$ as long as it falls off faster than $1/r$ at large $r$.

We will use the DeTurck method, which was first introduced in \cite{Headrick:2009pv} and further developed in \cite{Figueras:2011va}. For a recent review of this method, including some implementation details, see \cite{Dias:2015nua}. We seek solutions with a single (planar) horizon with $T>0$.  Our metric and gauge field \emph{ans\"atze} read
\begin{subequations}
\begin{multline}
\dd s^2 = \frac{L^2}{(1-y)^2}\Bigg\{-y\,y_+^2 g(y)\, q_1(x,y)\dd \tilde t^2+\frac{q_2(x,y)}{y\,g(y)}\left[\dd y+\frac{y\,q_3(x,y)\dd x}{(1-x^2)^2}\right]^2 \\ +\frac{4\,y_+^2\,q_4(x,y)\,\dd x^2}{(2-x^2)(1-x^2)^4}+\frac{x^2(2-x^2)\,y_+^2\,q_5(x,y)}{(1-x^2)^2}\dd \phi^2\Bigg\}
\label{eq:line}
\end{multline}
and
\begin{equation}
A = L\, y\,q_6(x,y)\,\dd \tilde t\,,
\end{equation}
\end{subequations}
where $x$ and $y$ each take values in $(0,1)$, $g(y)=3-3y+y^2$, and $q_1,\ldots,q_6$ are unknown functions of $x$ and $y$. In these coordinates, $x=0$ is the axis, $x=1$ is the asymptotic region infinitely far from the axis, $y=0$ is the horizon, and $y=1$ the location of the conformal boundary. The DeTurck method requires a reference metric which we choose to be the line element (\ref{eq:line}) with $q_1 = q_2 = q_4 = q_5 =1$ and $q_3=0$, which corresponds to the planar Schwarzschild black hole.  We have also redefined the time coordinate $t=L\tilde t$ so that the AdS length $L$ sets a scale and does not appear in the equations of motion. 

We will consider solutions with temperature $T>0$.  At the moment, we have three parameters given by the amplitude $a$, the profile width $\sigma$, and the temperature $T$. Conformal symmetry implies that only the conformally invariant products $a\,\sigma$ and $T\,\sigma$ are physical. Therefore we can set $\sigma = 1$ without any loss of generality.

At the boundary, located at $y=1$, there is a simple relation between $x$ and the radial coordinate of (\ref{eq:bndisflat})
\begin{equation}
r=\frac{x\sqrt{2-x^2}}{1-x^2}\,.
\end{equation}
This implies that the chemical potential $\mu$ in the $x$ coordinates takes the form
\begin{equation}
\mu(x) = a\,(1-x^2)^8\,,
\end{equation}
where have used conformal freedom to set $\sigma = 1$.  As boundary conditions here, we must demand that the metric is the same as the reference metric, and that $q_6(x,1)=\mu(x)$.  We must also impose similar boundary conditions at the other asymptotic infinity at $x=1$.  The metric must be the same as the reference metric and $q_6(1,y)=0$.

The horizon in (\ref{eq:line}) is located at $y=0$ and has an associated Hawking temperature
\begin{equation}
T = \frac{3\,y_+}{4\pi\,L}\,
\end{equation}
which is the temperature given by the reference metric.  We thus see that $y_+$ appearing in (\ref{eq:line}) essentially parametrises  the temperature of the horizon. The boundary conditions here and at the axis $x=0$ are determined by regularity.  In the DeTurck method, these conditions are automatically imposed by our choice of ansatz and reference metric if the functions remain finite, which is always true while performing numerics. For more details on boundary conditions, we refer the reader to \cite{Horowitz:2014gva,Dias:2015nua}.

In order to solve this problem numerically, we use a standard Newton-Raphson algorithm and discretise the resulting linear equations using Chebyshev grids in each of the directions $x$ and $y$. A seed is readily provided by the planar Schwarzschild-AdS solution when $a=0$. We can then explore the regime of interest where $T$ is small, and $a$ is large. At these extreme parameters, there are large gradients in our functions $q_i$. In order to resolve these gradients, we use at least $200$ points in each of the directions. Since we are using Chebyshev grids, this necessarily leads to large, dense, and poorly conditioned matrices. Therefore, to solve the linear system at each Newton iteration, we opt to use an iterative Krylov solver with a finite difference preconditioner rather than a direct LU decomposition.  At extremely low temperatures, we resort to LU decomposition with octuple precision.  

\section{Results}

\subsection{Black Mushrooms}

\begin{figure}
\centering
\subfigure[\,Various $a$ at constant $T=0.119366$.]{
\includegraphics[height=0.39\textwidth]{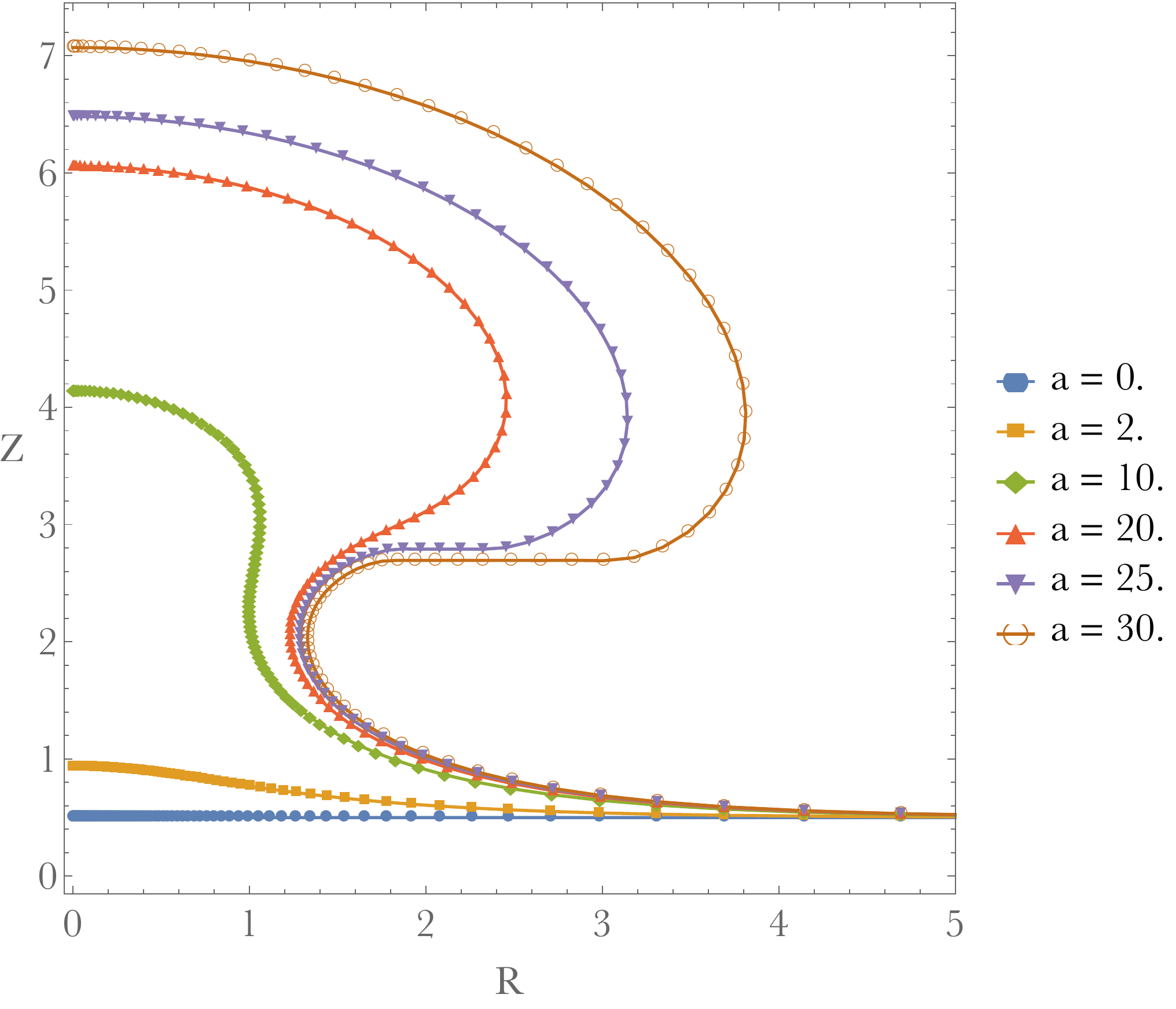}
\label{figs:isoa}
}
\subfigure[\,Various $T$ at constant $a=10$.]{
\includegraphics[height=0.39\textwidth]{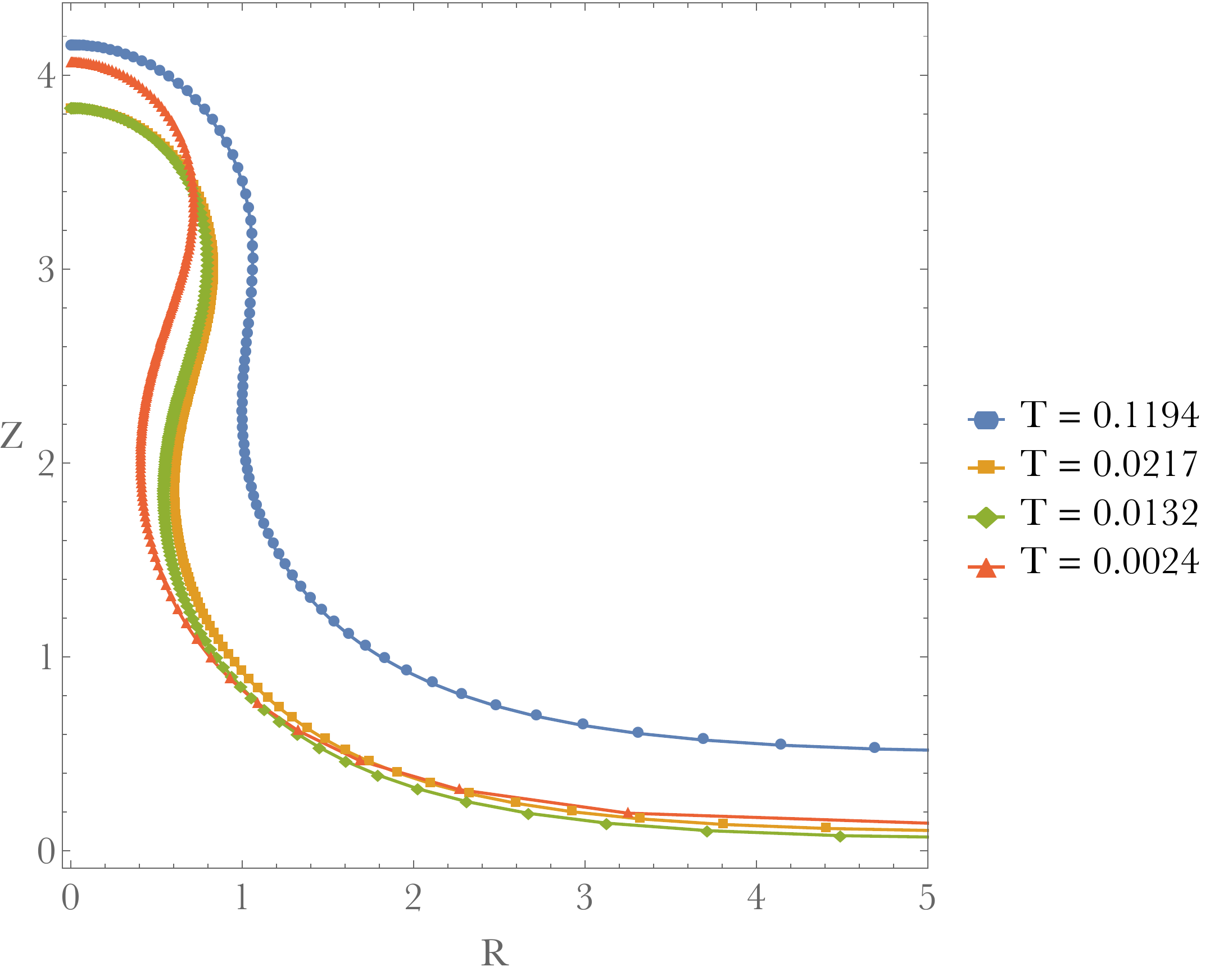}
\label{figs:isob}
}
\caption{\label{figs:iso} Isometric embeddings of the deformed planar horizon into $\mathbb{R}^3$.}
\end{figure}

\begin{figure}[ht]
\centering
\includegraphics[width=0.5\textwidth]{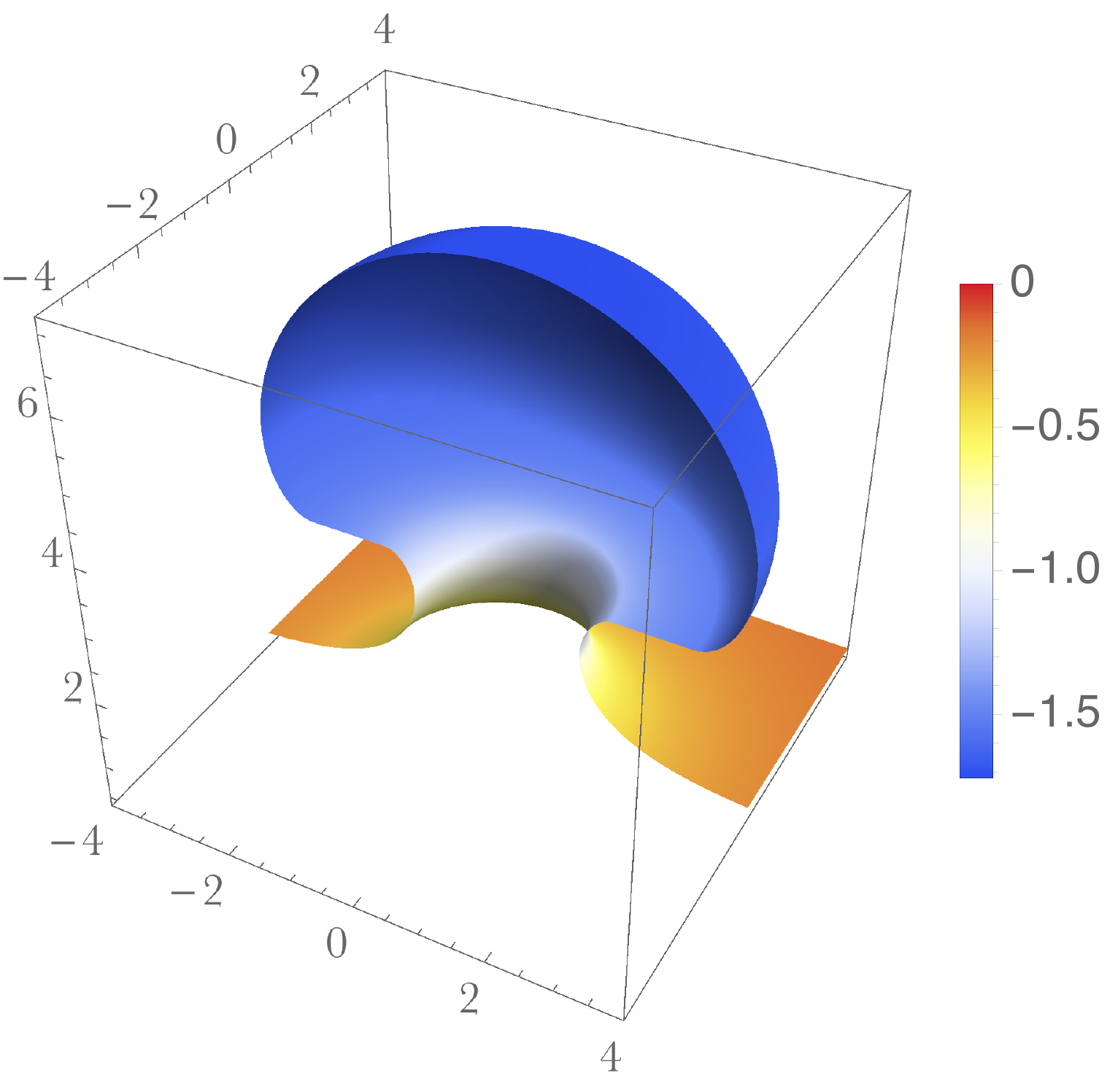}
\caption{\label{fig:iso3d} Isometric embedding of the deformed planar horizon into $\mathbb{R}^3$, coloured with the charge density on the horizon.  This solution has $a=30$ and $T=0.119$.}
\end{figure}

By construction, at $a=0$, our solutions are the same as that of planar Schwarzschild-AdS.  However, as one increases the amplitude $a$, the electric field near the axis becomes stronger, and the planar horizon becomes deformed. This family seems to exist for all $a$ and $T>0$.  For large $a$ and small $T$, the horizon becomes so highly deformed, as to resemble the shape of a mushroom.  This can be seen in  Figs.~\ref{figs:iso} and \ref{fig:iso3d}, where we present isometric embeddings of the induced horizon geometry into $\mathbb R^3$.  (For convenience, we make all of our plots with $L=1$.) As seen in the embedding Fig.~\ref{fig:iso3d}, the `cap' of the mushroom contains a higher charge density than the rest of the horizon geometry. It is as if a hovering black hole is trying to be pulled out of the planar horizon. 

\begin{figure}[ht]
\centering
\includegraphics[width=0.5\textwidth]{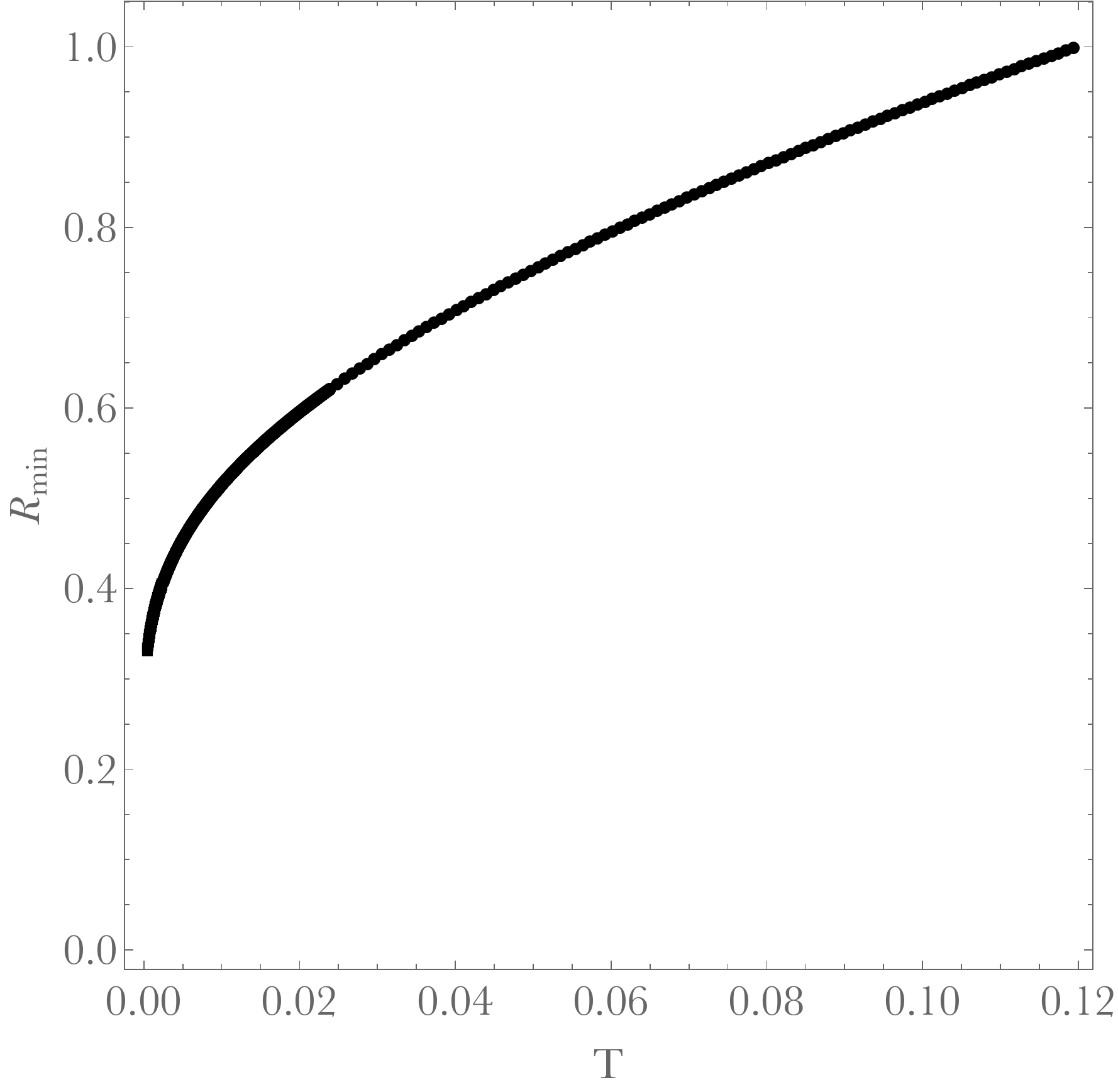}
\caption{\label{fig:rmin} The minimum radius of the neck region of the horizon as a function of temperature, for amplitude $a = 10$.}
\end{figure}

We would like to know if this horizon can pinch off to actually form a hovering black hole. Such a pinch-off will necessitate the formation of a naked singularity. To determine this, we study the minimum radius, $R_{\min}$, of the neck region of the horizon, between its mushroom cap and its asymptotic planar region (the minimum radius of the `stem'). Note that such a minimum radius is only well-defined for a sufficiently deformed planar horizon.  If we define a black mushroom by the existence of this feature, it seems that black mushrooms exist for all $a\gtrsim a_{\max}$ at low temperature, where $a_{\max}$ is the maximum amplitude for a nonsingular $T=0$ solution (without hovering black holes).  Fig.~\ref{fig:rmin} shows the minimum radius of this neck region as a function of temperature for $a  = 10 > a_{\max}$.  As one lowers $T$, this radius shrinks slowly at first, and then much more rapidly as $T \to 0$.   Fig.~\ref{fig:rmin} suggests that black mushrooms will not pinch off at nonzero temperature. In the following subsection, we will attempt an extrapolation to zero temperature.  

Let us also examine the total charge of the solution.  An argument was presented in \cite{Horowitz:2014gva} from the dual field theory side that the total charge must vanish at zero temperature.  This assumed that the solution was built up by slowly increasing the amplitude $a$, as we do in our proposed counterexample to cosmic censorship. This argument does not refer to the bulk and implies that any potential endpoint to our proposed counterexample must have vanishing total charge. In the classical limit, the endpoint might be singular.

\begin{figure}
\centering
\subfigure[\,$a=2$.]{
\includegraphics[height=0.46\textwidth]{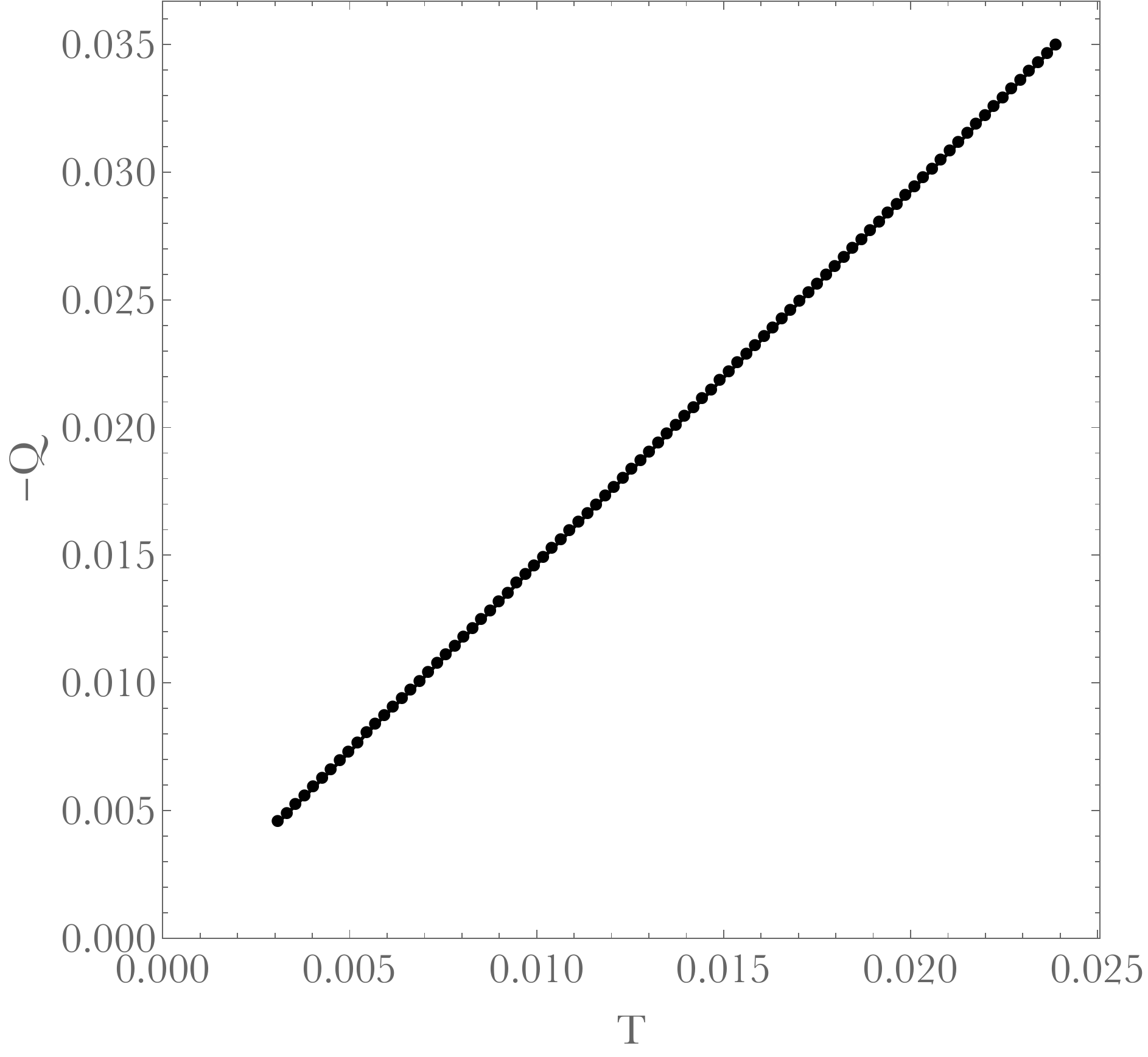}
\label{figs:isoa}
}
\subfigure[\,$a=10$.]{
\includegraphics[height=0.46\textwidth]{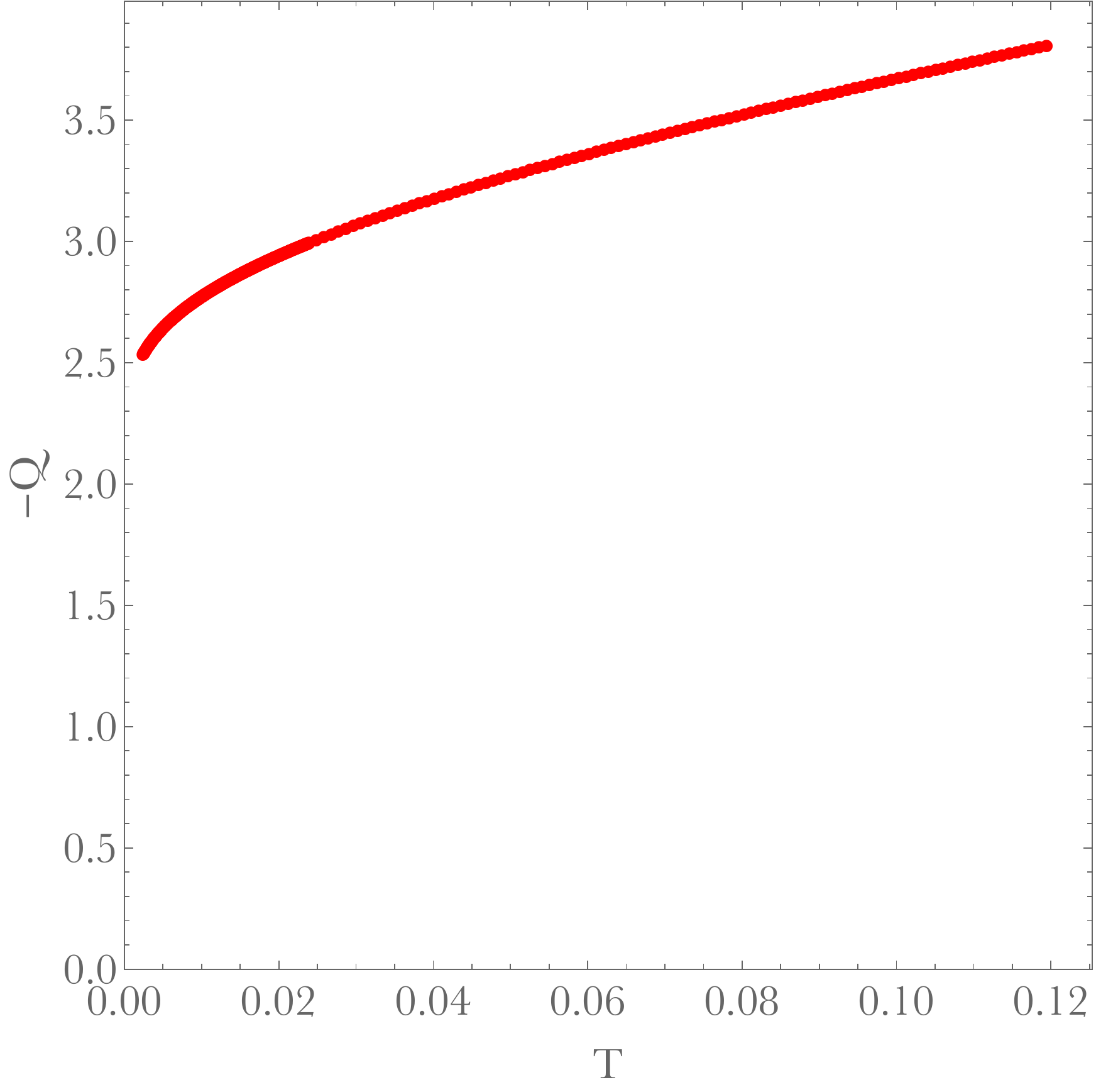}
\label{figs:isob}
}
\caption{\label{figs:char} Plots of the total charge as a function of temperature, for two values of the amplitude $a$.  For our profile, $a_{\max}\approx8$.}
\end{figure}

In Fig.~\ref{figs:char}, we show the total charge as a function of the temperature for two values of $a$: one below $a_{\max}$ and one above.  When $T>0$, both cases have nonzero total charge.  However, the total charge for $a<a_{\max}$ approaches zero charge linearly as $T\to0$, while the total charge for $a>a_{\max}$ decreases much more slowly. This behaviour for $a>a_{\max}$ appears similar to that of $R_{\min}$ in Fig.~\ref{fig:rmin}.


\begin{figure}[ht]
\centering
\includegraphics[width=0.5\textwidth]{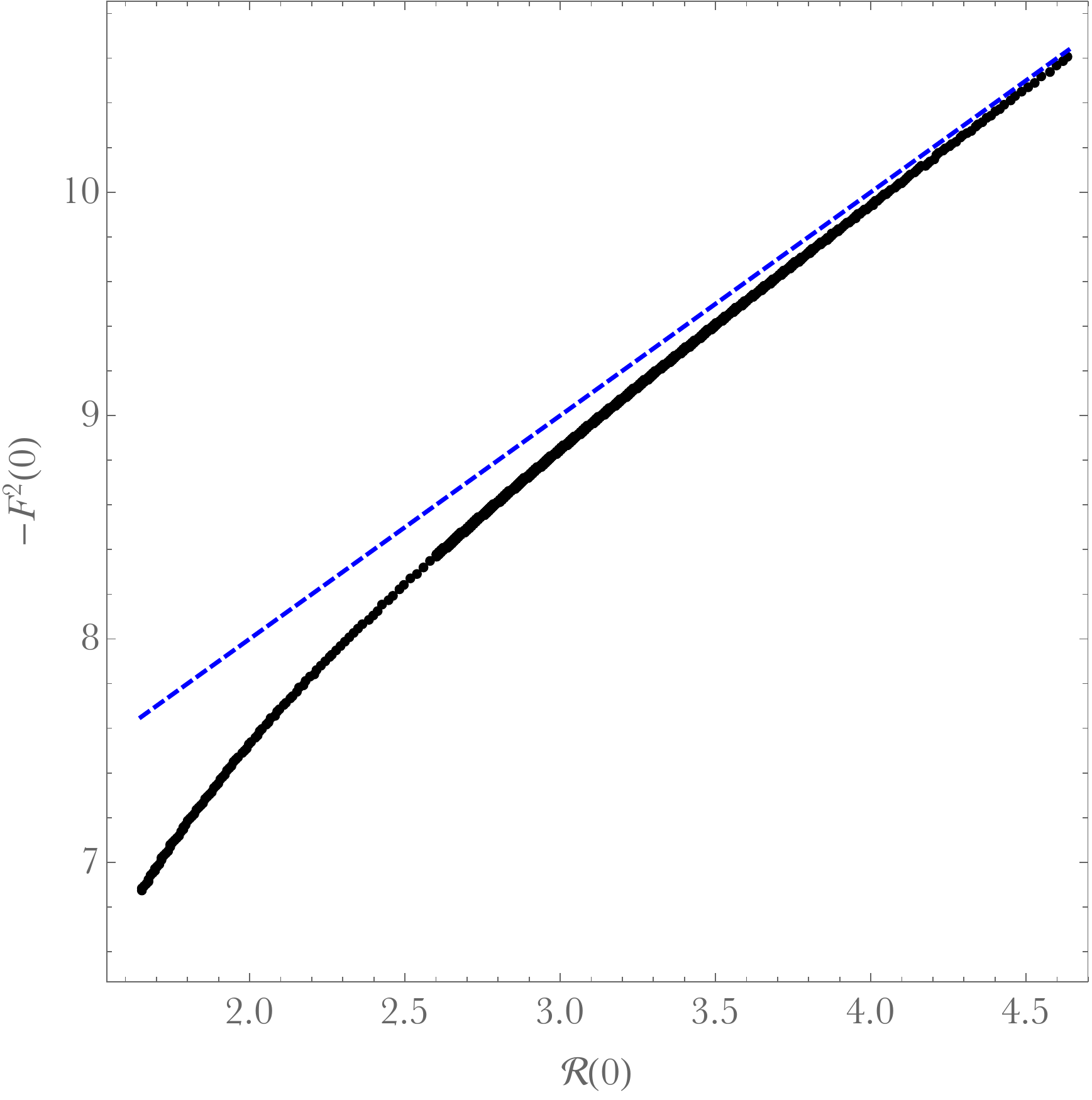}
\caption{\label{fig:rn} The square of the Maxwell field, $-F^2(0)$, and the induced scalar curvature of the horizon, $\mathcal R(0)$, both evaluated where the horizon meets the axis. The black dots represent black mushrooms at fixed amplitude $a = 10$ and different temperatures. The dashed blue line represents the same quantities for spherical extremal Reissner-Nordstr\"om black holes.  The two curves approach each other for low temperature black mushrooms.}
\end{figure}

The difference in behaviour between $a<a_{\max}$ and $a>a_{\max}$ as shown in Fig.~\ref{figs:char} might be explained by a difference in emergent IR geometries. Below $a_{\max}$, the zero temperature limit has a Poincar\'e horizon (an $AdS_4$ region). This was confirmed in \cite{Horowitz:2014gva} where the zero temperature solution was constructed.  But above $a_{\max}$, we may instead have an extremal horizon with a region resembling $AdS_2$.  To see if this is indeed the case, we can compare the top of the black mushrooms (where the horizon meets the axis) to a spherical extremal Reissner-Nordstr\"om black hole. Fig.~\ref{fig:rn} shows the square of the field strength $F^2\equiv F^{ab}F_{ab}$ on the horizon and the induced scalar curvature of the horizon $\mathcal R$,  both evaluated at the top of the black mushroom.  As the temperature of the black mushroom decreases, these quantities approach that of extremal Reissner-Nordstr\"om.

\subsection{Extrapolation to Zero Temperature}

At $T=0$ for our profile \eqref{eq:profile}, the results of \cite{Horowitz:2014gva} indicate that 
a curvature singularity appears at $a_{\max}\approx8$.  We would like to see if this is an isolated singularity at a particular value of $a_{\max}$, or if the singularity persists for larger values of $a>a_{\max}$.  More precisely, we would like to determine if $R_{\min}$ vanishes as $T\rightarrow 0$, which would imply a naked singularity. 

With our numerical capabilities, we have reached a temperature of approximately $2.4\times 10^{-4}$.  Reaching lower temperatures requires ever increasing numerical resources and computation time.  At some point, extrapolation will need to be used to infer the behaviour in the $T\rightarrow 0$ limit.   

Let us motivate the kind of extrapolation we will perform.  As we have seen, Fig.~\ref{figs:char} and Fig.~\ref{fig:rn} provide evidence that below $a_{\max}$, the emergent IR geometry is $AdS_4$, while above $a_{\max}$, there is an emergent $AdS_2$.  This implies that the low temperature physics of $a<a_{\max}$ solutions should be governed by $(2+1)$-dimensional conformal invariance, while that of $a>a_{\max}$ should be governed by $(0+1)$-dimensional conformal invariance. For the latter, the low-temperature physics can scale like a small power of the temperature \cite{Hartnoll:2012rj}. We would like to see if we have reached such a scaling regime by performing fits to our data. 

Let us set  $\beta = 1/T$ and make the following ansatz:
\be
R_{\min}(\beta) = c_1 \beta^{b_1} e^{c_2 \beta^{b_2}} \;.
\ee
We will find that both exponents, $b_i$ are negative so that for small $T$ or large $\beta$, this is equivalent to the form
\be
R_{\min}(\beta) = c_1 T^{-b_1}(1+c_2 T^{-b_2}+\ldots)\;,
\ee
which describes a scaling regime together with a correction term.  

The constant $b_2$ can by obtained by considering  
\be
f(\beta) \equiv \beta \frac{d}{d\beta} \log R_{\min} = b_1 + b_2 c_2 \beta^{b_2}\;.
\ee
Then the exponent $b_2$ can be computed via
\be
b_2 =  \beta   \frac{f''}{f'} +1\;,
\ee
where a prime denotes derivative with respect to $\beta$. A computation of the right hand side shows $b_2 = -1/3$ to high accuracy at low temperature. To be more precise, our numerical data deviates from $-1/3$ with a standard deviation of about $10^{-6}$.

\begin{figure}
\centering
\subfigure[\,Minimum radius $R_{\min}$.]{
\includegraphics[height=0.4\textwidth]{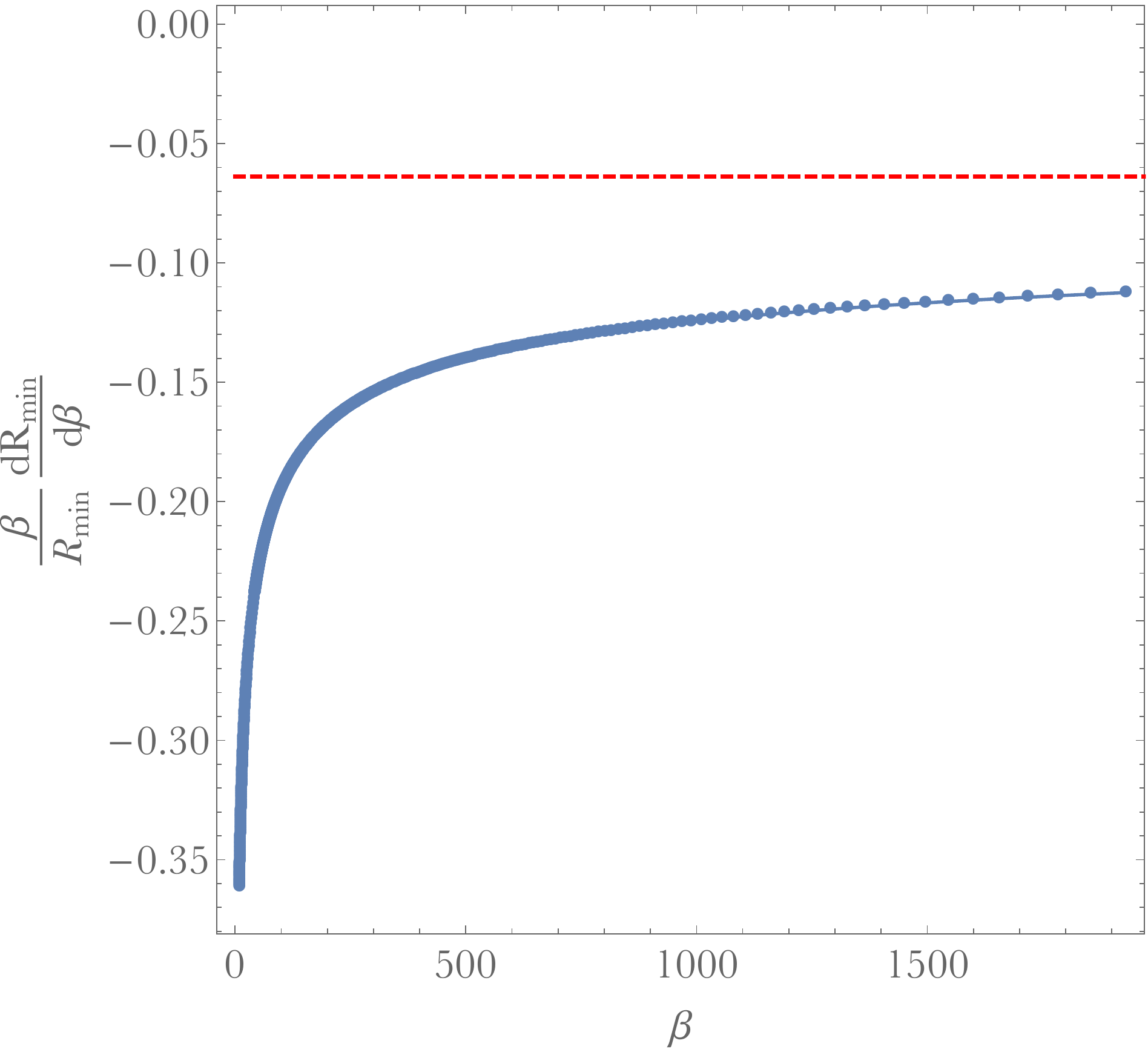}
\label{fig:rminfit}
}
\subfigure[\,Total charge $Q$.]{
\includegraphics[height=0.4\textwidth]{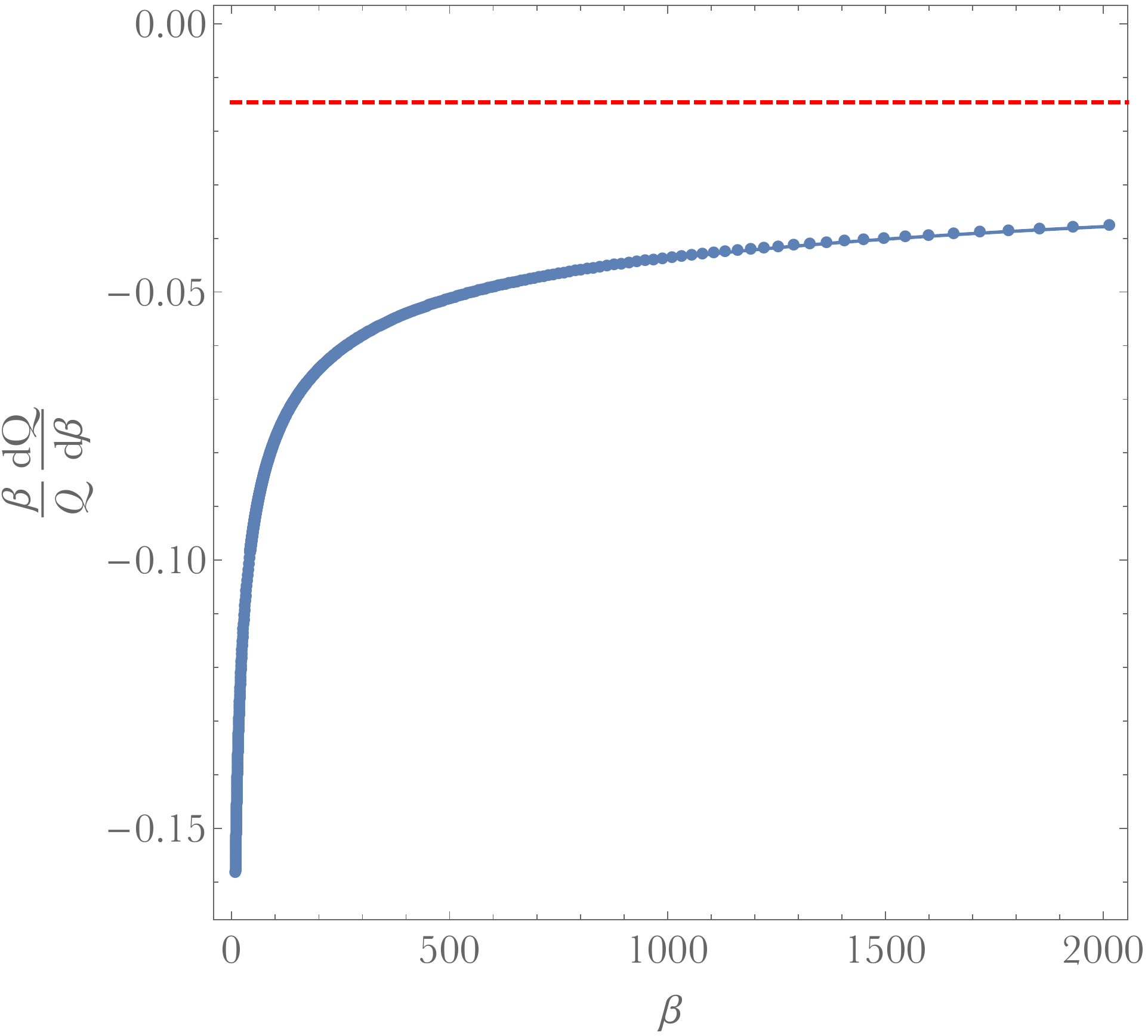}
\label{fig:Qfit}
}
\caption{\label{figs:fits} The logarithmic derivative of minimum radius $R_{\min}$ and total charge $Q$ with respect to $\beta = 1/T$. The dots are computed from our numerical solutions and the blue line is a fit to the form \eqref{eq:rminfit}. The red line denotes the value of $b_1$, and since it is negative for both fits, $R_{\min} \to 0$ and $Q \to 0$ as $\beta \to \infty$.}
\end{figure}

We next consider 
\be\label{eq:rminfit}
\beta \frac{d}{d\beta} \log R_{\min} = b_1 - \frac{1}{3}  c_2 \beta^{-1/3}
\ee
and perform a linear two-parameter fit for $b_1$ and $c_2$. The result is $b_1 = -.063$ and $c_2 = 1.84$. As shown in Fig.~\ref{fig:rminfit} the agreement is excellent. We have repeated this analysis at larger values of $a$ with similar results.
  
If we have a reached a scaling regime, other quantities should be governed by a similar scaling.  To study the low temperature behaviour of the charge, we do a fit in terms of $\beta = 1/T$ directly analogous to what we did for $R_{\min}$ above. The computation of $b_2$ for this case again gave us $b_2 = -1/3$ to high accuracy (the standard deviation from $-1/3$ being now $10^{-4}$). As before, we now perform a fit for the remaining parameters. The result is 
\be
 Q(\beta) = 1.239\ \beta^{-0.01567} \ e^{0.8576 \beta^{-1/3}} 
 \ee
which indicates that $Q$ will indeed vanish at $T=0$. Fig.~\ref{fig:Qfit} shows the excellent agreement of this fit with the data. Remarkably, the power of $\beta$ for $Q$ is precisely $1/4$ the power of $\beta$ for $R_{\min}$ (to four significant digits). This weak scaling with temperature is likely the result of IR conformal invariance.

The fact that the fits are of high quality and both $R_{\min}$ and the total charge have similar exponents suggests that we are near a scaling regime. The difference of $1/4$ could be explained by a different scaling dimension between $R_{\min}$ and $Q$. 
The fact that $b_1$ is negative for both fits suggests that the horizon is going to pinch off at zero temperature, but not before; and that the total charge will vanish at zero temperature.

\subsection{Non-uniqueness and the Phase Diagram}
While we have mostly been motivated by a possible violation of cosmic censorship, our results also have implications for the phase diagram of localised charged defect solutions.  Recall that at zero temperature, there was a point $a_*$ above which hovering black holes exist.  Let us now describe what happens to this critical point as we move away from $T=0$.  In \cite{Horowitz:2014gva}, $a_*$ was obtained by asking when static orbits exist for charged test particles, and we can use the same criterion for $T>0$.\footnote{For the $T=0$ case in \cite{Horowitz:2014gva}, we used charged particles with $q=m$.  This criteria  continues to hold for $T>0$ since keeping $T$ fixed while reducing the mass of a Reissner-N\"ordstrom black hole still gives $q=m$ to leading order.}  As in \cite{Horowitz:2014gva}, we assume that there is no potential difference between the planar horizon and the hovering black hole,  i.e., we can set $A_t =0$ at both horizons.   Then hovering black holes exist if the effective potential for $q=m$ test particles
\begin{equation}
\mathcal{V} = \sqrt{-g_{tt}}-A_t\, 
\label{eq:potgeo}
\end{equation}
 has an extremum below zero.  
 
\begin{figure}[ht]
\centering
\includegraphics[width=0.67\textwidth]{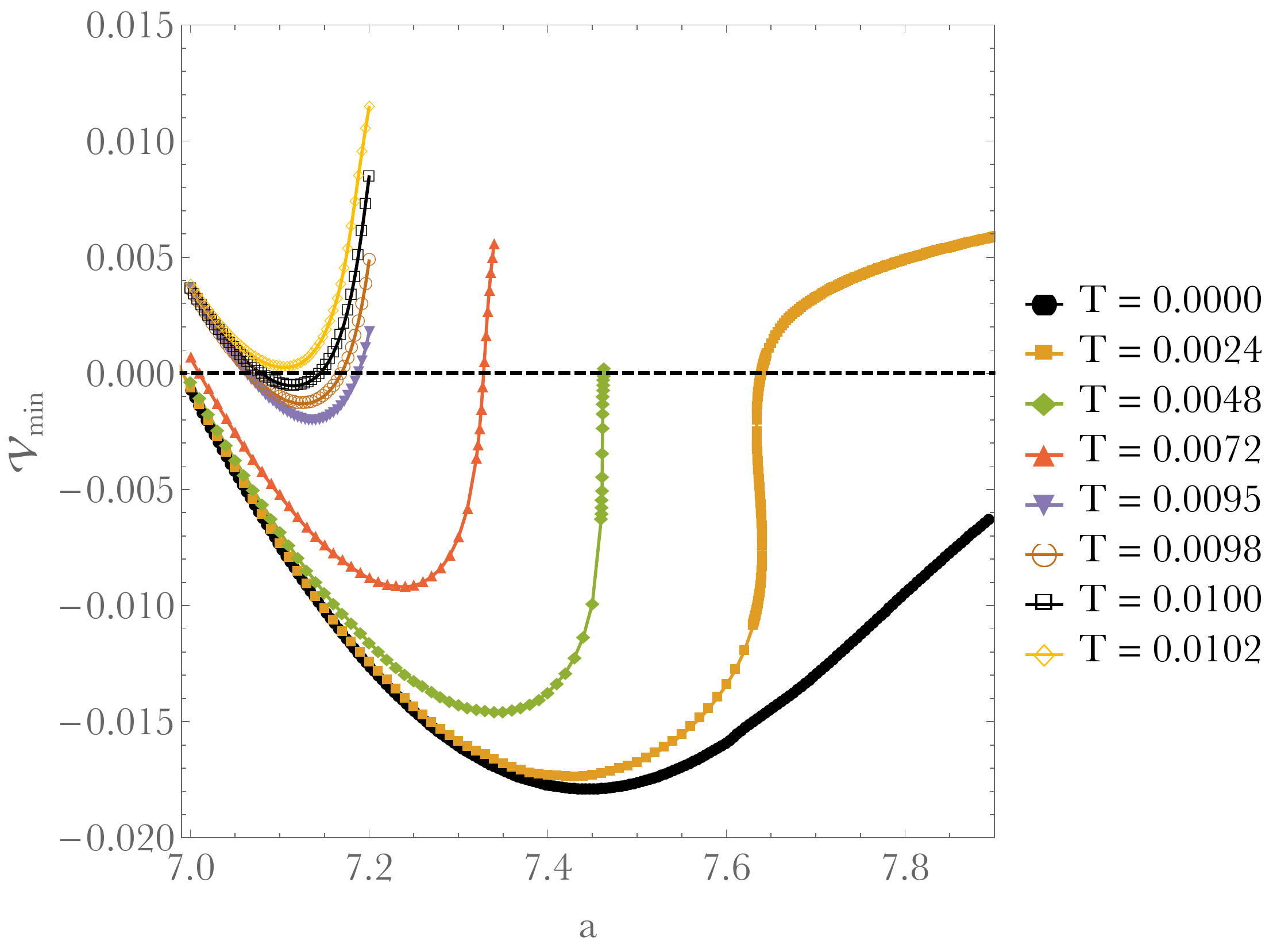}
\caption{\label{fig:orbi}Minimum of the effective potential for charged geodesics as a function of $a$, plotted for several low temperatures. The lowest black curve is for zero temperature.  Hovering black holes appear if $\mathcal V_{\mathrm{min}}<0$.}
\end{figure}

\begin{figure}
\centering
\subfigure{
\includegraphics[height=0.4\textwidth]{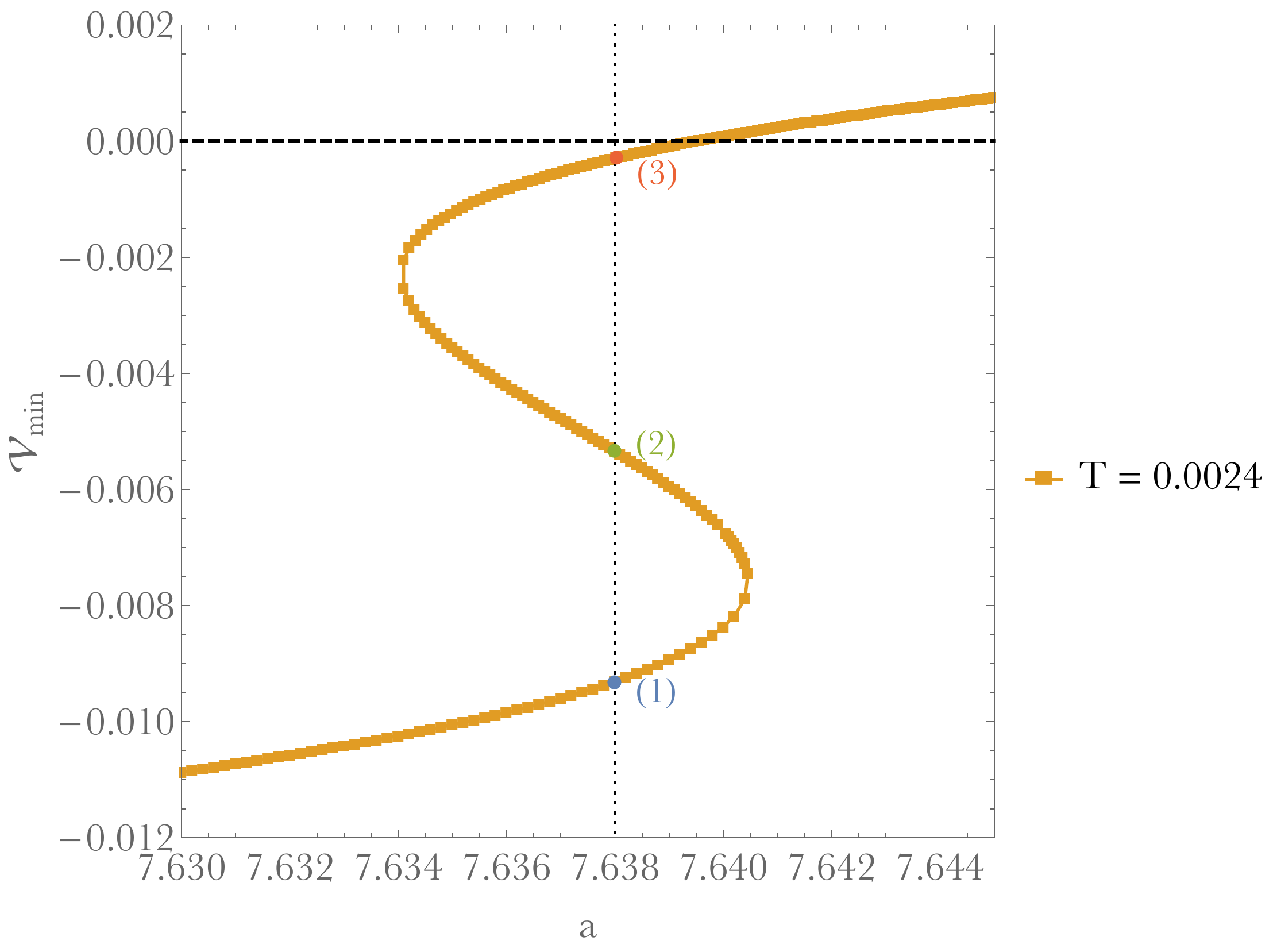}
\label{figs:isoa}
}
\subfigure{
\includegraphics[height=0.4\textwidth]{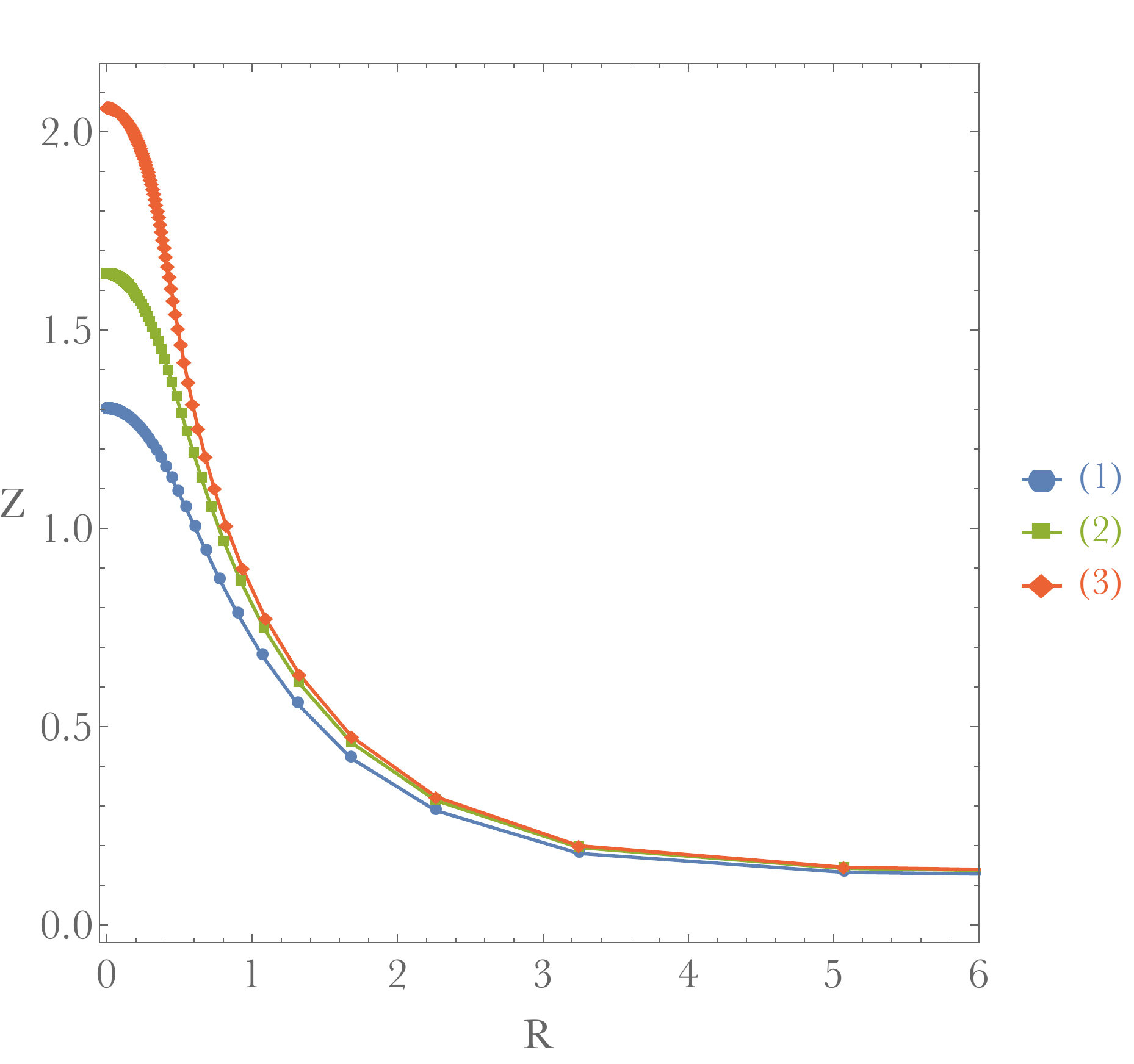}
\label{figs:isob}
}
\caption{\label{figs:nonunique} Left: A blow-up of the $T=.0024$ curve in Fig.~\ref{fig:orbi} showing three different solutions for the same $a$ and $T$.  Right: An embedding diagram of the horizons for the three solutions with the same $a$ and $T$.}
\end{figure}

  In Fig.~\ref{fig:orbi}, we show the minimum value of this potential $\mathcal V_{\mathrm{min}}$ as a function of $a$ for various temperatures.  Points where $\mathcal V_{\mathrm{min}}=0$ correspond to a phase boundary where small hovering black holes appear. We see that as one lowers the temperature, there is a critical value $T_c\approx0.01$ below which small hovering black holes can exist.  Surprisingly, a close examination of this plot shows that solutions are not uniquely labeled by $T$ and $a$. The plot on the left in Fig.~\ref{figs:nonunique} shows a blow-up of the $T=.0024$ curve indicating three different solutions for the same $a,T$. The right hand side of Fig.~\ref{figs:nonunique} shows the embedding diagram of the horizon for these three solutions. Notice the extreme sensitivity of the solution to the value of the amplitude $a$ in this region. The three solutions only exist in a narrow range where $a$ changes by about $.1\%$, but the height of the horizon jumps by a factor of two over this range. The nonuniqueness appears to start below $T\approx .0048$. 
  
It is not uncommon to have more than one AdS black hole solution with the same temperature. Consider spherical Reissner-N\"ordstrom AdS black holes with a small charge $Q$. Since the temperature of these black holes behaves like Schwarzschild AdS when $M\gg Q$ and goes to zero in the extremal limit, there is a range of temperatures for which there exist three solutions with the same $T$. Similar nonuniqueness has been seen previously for planar black holes in AdS \cite{Horowitz:2014gva,Santos:2014yja}.

\begin{figure}[ht]
\centering
\includegraphics[width=0.5\textwidth]{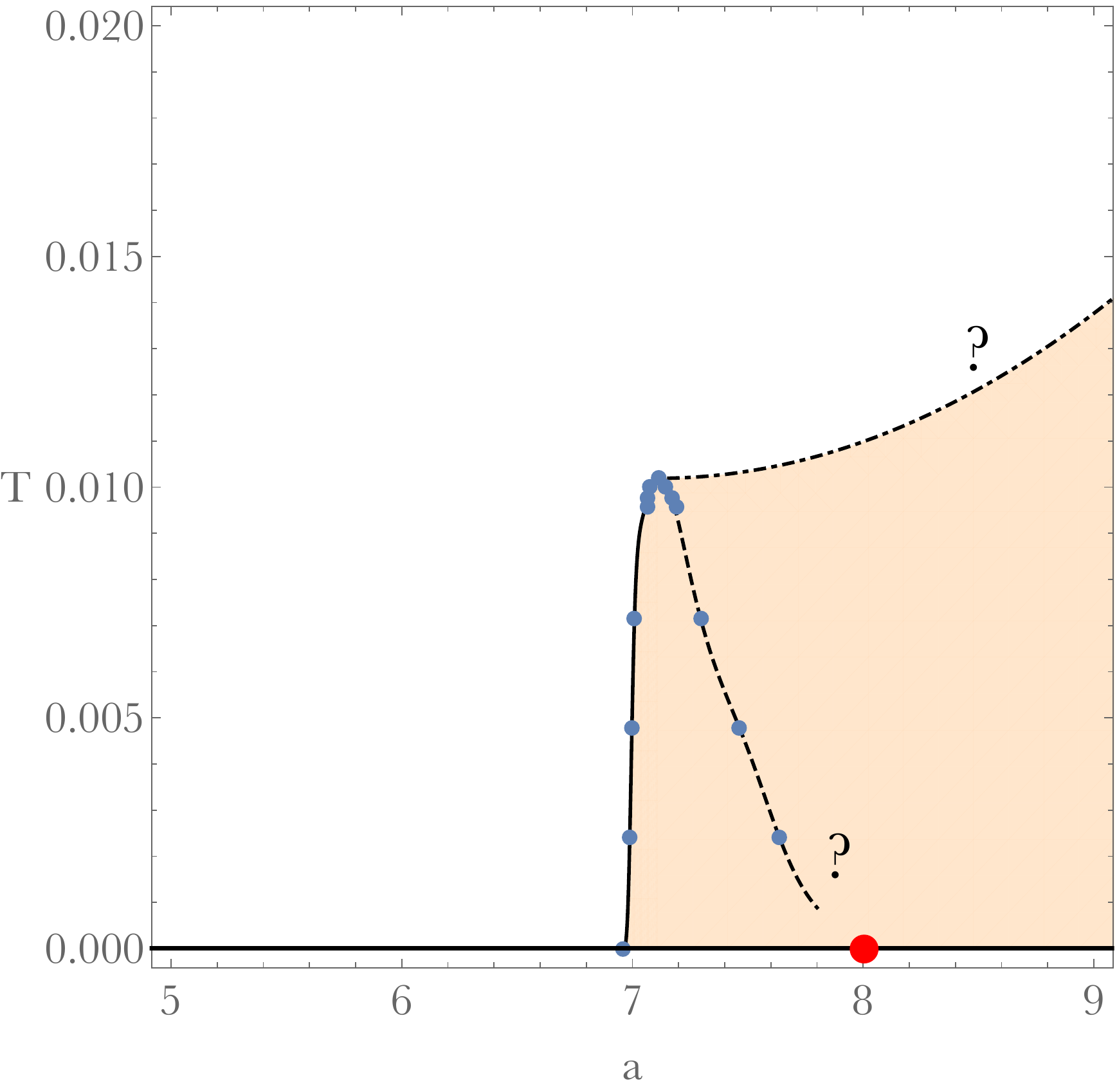}
\caption{\label{fig:phasediagram}Conjectured phase diagram of hovering black holes.  The small blue dots are numerical data indicating the appearance of small hovering black holes. The large red dot is $a_{\max}$. Hovering black holes are conjectured to exist within the shaded region.  The dashed line is the location of a phase boundary marking the appearance of a second family of small hovering black holes. The region to the right of this dashed line is conjectured to contain two hovering black hole families that compete.  For all $T>0$, there are also competing solutions without hovering black holes (not shown).}
\end{figure}

A possible  $T$ vs. $a$ phase diagram of hovering black holes is shown in Fig.~\ref{fig:phasediagram}.  From Fig.~\ref{fig:orbi}, it is clear that for each $T<T_c$, there are two values of $a$ where $\mathcal V_{\mathrm{min}}=0$. This  implies that small black holes exist along the dome shaped curve in Fig.~\ref{fig:phasediagram}. The data   suggests that the lower $a$ side of this curve connects with $a_*$ at zero temperature and higher $a$ side connects with  $a_{\mathrm{max}}\approx 8$.
Presumably, once a small hovering black hole can exist, it grows with increasing $a$ and fixed $T$. (This was the case at $T=0$.)   The dome shaped curve for small black holes  then implies  that there are two competing hovering black hole solutions for large $a$.   One of these families appears on the small $a$ side of the dome, and the other on the large $a$ side of the dome.  These two families compete with each other at still larger $a$.  Whenever these two families meet, there is a turning point in the solutions. 
Hovering black holes can also disappear by merging with the planar horizon. 

In addition to all these hovering black hole solutions, we have argued above that for all $a$ and $T>0$ there are also solutions without hovering black holes, so these would be a third competing phase in this picture.  The fact that this additional phase is not unique further complicates the phase diagram. We know little about which phases are dominant in the phase diagram, except that at $T=0$, the hovering black holes found in \cite{Horowitz:2014gva} are preferred over solutions without hovering black holes. We expect this will continue for $T>0$, but verifying this conjecture and finding the dominant phases will require explicitly constructing $T>0$ hovering black holes.
 
\section{Discussion}
We have investigated the space of static Einstein-Maxwell solutions with boundary condition $A_t |_\partial  = a\,p(r)$ (representing a charged defect on the boundary).  We chose a fixed profile $p(r)$ that decays rapidly at large $r$, and parametrised the solutions by the amplitude $a$ and the temperature $T$. Among these solutions, there are two types of known competing phases: one containing a hovering black hole and one without. 

Our results suggest that the family of solutions without hovering black holes contains a naked curvature singularity for parameters $T=0$ and $a>a_{\max}$.  For $T>0$, this singularity is not present, but the planar horizon can become highly deformed into a mushroom-like geometry.  We have also shown that for low $T$, and $a$ below but near  $a_{\max}$, the solutions are not uniquely characterised by $T$ and $a$.

We conjectured in the first section that the presence of the naked singularity leads to a violation of cosmic censorship in a dynamical scenario. The idea is to impose a boundary condition in which the amplitude $a$ increases slowly from zero to a constant value larger than $a_{\max}$. However, we have stopped short of carrying out the fully time-dependent simulation.

We further note that a field theory argument in \cite{Horowitz:2014gva} implies that any endpoint to such an evolution must have vanishing total charge. Our results suggest that solutions without hovering black holes indeed have zero total charge in the $T\rightarrow0$ limit, and are thus consistent with being an endpoint by this argument. 

It may be possible to also construct a counterexample to cosmic censorship using time independent boundary conditions (which conserve energy).  If one simply requires that the amplitude is a constant $a>a_{\max}$, then any smooth generic initial data with  a Poincar\'e horizon in the interior should evolve to form arbitrarily large curvature at late time.   The reason is the same as above: hovering black holes cannot form dynamically and the only known static endstate has a naked singularity. We leave the construction of such initial data and a study of its evolution to future work.

Our proposed counterexample to cosmic censorship uses the Maxwell field in an important way. It is natural to ask if similar counterexamples could be constructed using other matter fields.  If one considers gravity coupled to just a scalar field,  
there are a class of examples using ``designer gravity" \cite{Hertog:2004ns} which have the property that a finite change in the boundary conditions on the scalar field can cause the static solutions to become singular \cite{Hertog:2006rr}. 
 Another alternative is to consider pure gravity, and deform the boundary metric away from flat space to add angular momentum.  We find that naked singularities can form. However in both of these cases, it is unclear whether (neutral) hovering black holes can form in the dynamic evolution.  In the latter case, spin-spin interaction may keep the black hole from falling towards the Poincare horizon. If hovering black holes can form, then naked singularities do not arise. Note that this is unlike the case considered here with a Maxwell field, where hovering black holes cannot form in the dynamic evolution since they are charged and there is no charged matter to form them. 

We have made some preliminary comments about the phase diagram for hovering black holes. The results of \cite{Horowitz:2014gva} have shown that at $T=0$, hovering black holes exist for $a>a_*$, for some critical amplitude $a_*<a_{\max}$.  They seem to exist for arbitrarily large $a$.  Here, we have extended this critical amplitude to $T>0$.  Hovering black holes exist below some critical temperature $T_c$, but at least for some  range of temperature $T<T_c$, we find that there are two critical values of $a$ where small hovering black holes appear.  One of these connects to $a_*$ in the zero temperature limit, while the other appears to connect to $a_{\max}$.  Completing the phase diagram will require numerically constructing hovering black hole solutions with $T>0$.  Due to the presence of the planar horizon, the integration domain for such solutions is more complicated than that of the hovering black holes constructed in \cite{Horowitz:2014gva}.

There is a possible connection with our result and the weak gravity conjecture \cite{ArkaniHamed:2006dz}, which states that any consistent quantum theory of gravity must contain charged particles with $q\ge m$. The point is that if we add a charged scalar field to the bulk, then it is possible that the black mushrooms we have constructed become unstable to forming nonzero scalar ``hair"  outside the horizon at low temperature. This would be analogous to what happens in a holographic superconductor \cite{Gubser:2008px,Hartnoll:2008vx}.  If this happens, the charge on the black mushroom is reduced and the horizon might remain nonsingular all the way down to zero temperature. 
 Clearly as one increases $q/m$, it becomes easier to create the scalar hair, and there might be a threshold value  such that cosmic censorship is preserved only for  $q/m$ larger than this value. It would be interesting to investigate this further to see if such a threshold exists and agrees with that predicted by the weak gravity conjecture\footnote{We thank C. Vafa for a discussion about this possible connection.}.

Finally, let us comment on the field theory implications for this singularity. In \cite{Horowitz:2014gva}, it was argued from dimensional analysis that profiles that decay faster than $1/r$ are \emph{irrelevant} deformations and have an IR geometry that is the usual Poincar\'e horizon.   However, even though the profile we choose here falls off much faster than $1/r$, we are finding some solutions with a singular IR geometry.  This is indicative of a \emph{dangerously irrelevant} deformation, where a deformation that is irrelevant in the UV can later become relevant in the IR.  If cosmic censorship is violated in the bulk, this may be an indication that the field theory is poorly behaved without the addition of charged matter fields.  

\vskip 0.5cm
\centerline{\bf Acknowledgements}
\vskip .5 cm
We thank Sean Hartnoll, Donald Marolf, Harvey Reall, and Cumrun Vafa for discussions. G.H. was supported in part by NSF grants PHY-1504541. The research leading to these results has received funding from the European Research Council under the European Community's Seventh Framework Programme (FP7/2007-2013) / ERC grant agreement no. [247252].  B.W. is supported by European Research Council grant no. ERC-2011-StG 279363-HiDGR. This work was partially undertaken on the COSMOS Shared Memory system at DAMTP, University of Cambridge operated on behalf of the STFC DiRAC HPC Facility. This equipment is funded by BIS National E-infrastructure capital grant ST/J005673/1 and STFC grants ST/H008586/1, ST/K00333X/1.


\bibliographystyle{JHEP}
\bibliography{all}
  
\end{document}